\documentclass[preprint,12pt,3p]{elsarticle}
\graphicspath{ {./figures/} }
\usepackage{subcaption}

\usepackage{amsmath,amsfonts}

\usepackage{algorithm}
\usepackage{algcompatible}
\usepackage{hyperref}
\usepackage{float}
\usepackage[justification=centering]{caption}
\usepackage{verbatim} 
\usepackage{apalike}
\usepackage{array}
\usepackage{amssymb}
\usepackage[
singlelinecheck=false 
]{caption}
\usepackage[usenames,dvipsnames]{color}

\usepackage[english]{babel} 
\usepackage{booktabs}
\usepackage[utf8]{inputenc}
\biboptions{authoryear}
\usepackage[shortlabels]{enumitem}
\usepackage{makecell}
\usepackage{lipsum}
\usepackage{url}
\usepackage{graphicx}
\usepackage{tikz}
\usepackage{pgfplots}
\usepackage{multirow}
\usetikzlibrary{arrows,positioning,automata}
\usepackage{balance}
\usepackage{amsmath}
\usepackage[T1]{fontenc}
\usepackage{algorithm}
\usepackage{algpseudocode}
\usepackage{graphicx}
\usepackage[utf8]{inputenc}
\usepackage{balance}
\usepackage{color}
\usepackage{caption}
\usepackage{comment}
\usepackage{hyperref}
\usepackage{array}
\usepackage{soul}
\usepackage{color}
\usepackage{lipsum}
\usepackage{balance}
\usepackage{parskip}
\usepackage{longtable}

\newcommand{\eat}[1]{}

\pgfplotsset{compat=1.18}
\expandafter\def\csname ver@subfig.sty\endcsname{}
\usepackage{svg}
\usepackage{hyperref}
\begin{document}
\begin{frontmatter}



\title{A Human-Centered Approach for Improving Supervised Learning}
\author[label1]{Shubhi Bansal}
\ead{phd2001201007@iiti.ac.in}
\author[label1]{Atharva Tendulkar}
\ead{ms2104101010@iiti.ac.in} 
\author[label1]{Nagendra Kumar}
\ead{nagendra@iiti.ac.in}

\address[label1]{Department of Computer Science and Engineering,
Indian Institute of Technology, Indore, India}
\begin{abstract}
Supervised Learning is a way of developing Artificial Intelligence systems in which a computer algorithm is trained on labeled data inputs. Effectiveness of a Supervised Learning algorithm is determined by its performance on a given dataset for a particular problem. In case of Supervised Learning problems, Stacking Ensembles usually perform better than individual classifiers  due to their generalization ability. Stacking Ensembles combine predictions from multiple Machine Learning algorithms to make final predictions. Inspite of Stacking Ensembles’ superior performance, the overhead of Stacking Ensembles such as high cost, resources, time, and lack of explainability create challenges in real-life applications. This paper shows how we can strike a balance between performance, time, and resource constraints. Another goal of this research is to make Ensembles more explainable and intelligible using the Human-Centered approach. To achieve the aforementioned goals, we proposed a Human-Centered Behavior-inspired algorithm that streamlines the Ensemble Learning process while also reducing time, cost, and resource overhead, resulting in the superior performance of Supervised Learning in real-world applications. To demonstrate the effectiveness of our method, we perform our experiments on nine real-world datasets. Experimental results reveal that the proposed method satisfies our goals and outperforms the existing methods.
\end{abstract}
\begin{keyword}
Human-Centered AI \sep Supervised Learning \sep Ensemble Learning
\end{keyword}

\end{frontmatter}
\section{Introduction}
\label{sec:int}
Human-Centered AI is concerned with the realization that we are designing a system for users, as a result, it takes into account other important factors such as human psychology, design, ethics, constraints, business objectives, and explainable AI in addition to mathematics. Recently, Human-Centered Artificial Intelligence (AI) has grown significantly since Google announced its People $+$ AI Research programme in 2017 and Stanford University established its Institute for Human-Centered AI in 2019. The Human-Centered approach keeps humans informed so that they can oversee algorithmic decisions~\cite{riedl2019human,xu2019toward}. In Human-Centered AI, intelligent systems are built keeping in mind various real-life constraints~\cite{riedl2019human}. In this paper, we propose a Human-Centered approach to improve an important part of AI i.e,. Supervised Learning.

Supervised Learning is one of the most common methods in Machine Learning today \cite{rehman2023user, anshul2023multimodal,rehman2025context,bansal2024hybrid,bansal2022hybrid,bansal2024multilingual}. Supervised Learning is a way of developing Artificial Intelligence systems in which an algorithm is trained on input data that has been labelled for a specific output. In Supervised Learning problems, Ensemble Learning usually performs better than individual classifiers due to their generalization ability ~\cite{xiao2019svm,aljame2020ensemble,gao2019adaptive}. Instead of relying on a single model, Ensemble Learning combines predictions of multiple models. As an Ensemble Learning model is built by considering many base models,  it usually outperforms a single base model ~\cite{xiao2019svm,aljame2020ensemble,gao2019adaptive}. Ensemble Learning models can be broadly classified into three types, namely Bagging, Boosting, and Stacking~\cite{zhou2021ensemble}.

Despite the fact that Stacking Ensemble is a winning approach in Kaggle contests, the overhead of Stacking Ensemble, such as high cost, resources, time, and lack of explainability, poses issues in real-world Supervised Learning applications ~\cite{chen2006probabilistic,yan2018detecting,zhou2021ensemble}. Users typically spend a significant amount of time looking for the best possible combination of base models to improve performance without understanding the reason for the increase in performance ~\cite{xiao2019svm,aljame2020ensemble,gao2019adaptive}. Though some researchers claim that merging models with variety can help to improve performance, the definition of measuring variety is not well-defined in the true sense ~\cite{melville2003constructing,kuncheva2003measures}. Existing research works in this direction have shown different ways to measure diversity ~\cite{melville2003constructing,kuncheva2003measures}. We have identified three main drawbacks in the existing Stacking Ensemble approaches. The knowledge gaps are as follows: i) Inadequate comprehension of Ensembles' fundamental mechanism, i.e., the less explicable nature of ensembles ii) Improving performance at the expense of a significant amount of time and resources. iii) The diversity of base models is not stated in a clear, simple, easy-to-implement, and explainable manner. Even though Ensemble Learning increases the performance, it is not used in real-world applications due to the aforementioned drawbacks ~\cite{chen2006probabilistic,yan2018detecting,zhou2021ensemble}.

To bridge the knowledge gaps, we have proposed a novel Human-Centered Behavior-inspired Ensemble Learning approach that is explainable, efficient and easy to implement. We begin by contemplating how humans would solve a similar problem in daily lives. We then attempt to translate this line of logic to code by developing a solution that combines the concept of strong base models, various feature extraction and reduction techniques, class probability, and clustering. We find all strong base models that provided satisfactory performance. We then seek to identify a minimum subset of the obtained strong models that serve as base models for Ensemble and aid in the improvement of Supervised Learning performance. Finally, after incorporating the results of the selected base models, the final predictions are obtained. In addition, using a Human-Centered approach, we ensure that the entire process is more comprehensible. The proposed solution will help to remove the black-box around Ensemble Learning i.e., how performance improves when certain base models are combined, thus making Ensemble Learning more explainable. Furthermore, time and computational overhead can be reduced without sacrificing much on performance, thereby assisting in the achievement of a balance between performance and time. As a result, Ensemble Learning can be employed to make Supervised Learning applications explainable and efficient.

Our key contributions are as follows:
\begin{itemize}
    \item We have presented a novel Human-Centered and explainable approach to improve Supervised Learning.
    \item We have presented a robust solution to efficiently utilize Ensemble Learning for Supervised Learning. 
    \item We have presented an efficient method for identifying diverse models that will serve as strong models for Ensemble Learning.
    \item The proposed method is designed by keeping time, resources, explainability and Human-Centered design constraints in mind.
    \item Experimental results show that our proposed method outperforms the existing state-of-the-art methods. 
\end{itemize}

The rest of the paper is organized as follows. Section~\ref{sec:rw} discusses the related works on Ensembles. In Section~\ref{sec:meth}, we present the overall design and discuss the proposed methodology. Section~\ref{sec:expevl} showcases the experiments and evaluates the performance of the proposed model. Finally, in Section~\ref{sec:cnf}, we conclude the article.

\section{Related Work}
\label{sec:rw}
This section reviews the related work. 
Designing AI with people in mind is crucial for societal progress and the common good \cite{raghaw2024explainable,dar2024contrastive,}. In~\cite{auernhammer2020human}, many philosophical stances are described, along with a number of Human-centered Design strategies, and their impact on the advancement of AI is discussed. The article discusses that humanistic design research should be a key component of cross-disciplinary partnerships with technologists and decision-makers. 
In ~\cite{wagstaff2012machine}, the author identifies six key challenges in current Machine Learning research that must be addressed in order to solve problems in a Human-Centered manner on a large scale and have a larger impact. Keeping in mind humans' demands and real-world constraints, we should build algorithms ~\cite{riedl2019human}. According to ~\cite{riedl2019human}, one way to incorporate Human-Centered behavior is to take inspiration from how individuals perform similar jobs.~\cite{riedl2019human} states that while ``Human-Centered AI'' does not imply that an AI or Machine Learning algorithm must think like a human, it does imply that humans should be able to understand algorithmic conclusions and rely on the recommended technique.

Stacking Ensemble typically outperforms their individual classifiers owing to its ability to generalize ~\cite{xiao2019svm,gao2019adaptive,reddy2020ensemble,rustam2020predicting}. Stacking Ensemble is used to increase performance on certain datasets.
The author ~\cite{chen2006probabilistic} employed Decision tree, AdaBoost, and Multi-tree ensembles to achieve an accuracy of 85.2\% on the NSL-KDD dataset.
In ~\cite{reddy2020ensemble}, the highest performing base model was enhanced from 77\% to 81.9\% by stacking Random Forest, Decision Tree, AdaBoost, KNN, and Logistic Regression. Using Stacking Ensemble, which included Random Forest, Gradient Boosting classifier, and Extra tree classifier, improved performance from 97.1\% to 98.2\% percent in ~\cite{rustam2020predicting}.
In ~\cite{kaur2022ensemble}, the authors employ an ensemble strategy to categorize benign and malignant brain MRI scans. The hybrid aspects of the MRI are retrieved, including its shape, intensity, color, and textural features. Separate classifiers using decision trees, k-nearest neighbor, and support vector machines are used on the feature set. The stacking model is then used to integrate each classifier's prediction and outcome to produce the final result.

Despite improved performance and ranking first in the Kaggle competition, one of the most significant challenges with the Stacking Ensemble is its lack of applicability in real-world applications. According to ~\cite{zhou2021ensemble}, the key limitations of Ensemble Learning are overhead of cost, resources, time, and lack of explainability. ~\cite{yan2018detecting,al2017successful,martinez2019ensemble} depict winning solutions in various Kaggle competitions. The Stacking Ensemble method suggested in~\cite{yan2018detecting} increased accuracy from 99.36\% to 99.83\%, and prediction time increased exponentially from 0.03 to 13.33 seconds. In~\cite{zhou2021ensemble}, the author goes in-depth about various Ensemble Learning techniques. The author further discusses the benefits and a few important constraints that explain the reason behind its infrequent usage in real-world applications.

In Stacking Ensemble, it is hypothesized that taking more diverse models will result in improved performance. Hence, many authors have conducted research on finding diversity among base models. In~\cite{melville2003constructing}, diversity among base models is obtained by taking artificial training examples. \cite{kuncheva2003measures} examines ten alternative techniques to quantify model diversity and demonstrates a relationship between diversity and performance. To overcome the overhead of time and resources, several approaches have been proposed in ~\cite{chen2006probabilistic,dai2013competitive,martinez2019ensemble} to find a minimum subset of base models with maximum diversity. ~\cite{chen2006probabilistic} uses the concept of probability, ~\cite{dai2013competitive} uses cross-validation technique while ~\cite{martinez2019ensemble} uses the concept of margin maximization to find minimal base models.
AlJame et al.~\cite{aljame2020ensemble} created an ensemble learning model called ERLX to diagnose COVID-19. By utilizing two levels of classifiers, the suggested model made use of structural diversity. Extra trees, Random Forests, and Logistic Regression were used to feed the first-level classifiers' predictions into the XGBoost second-level classifier to improve prediction performance. In~\cite{du2022study}, the authors develop a feature engineering-based ensemble learning prediction strategy for use in academic settings that fully exploits the benefits of random forest for feature extraction and the predictive power of XGBoost. The storage of individual forecasts and the pricey retraining of all candidate models for an ensemble, which are frequently needed by previous similar systems, are avoided by a novel approach to ensemble selection in~\cite{kazmaier2022power}.

In this paper, we propose a Human-Centered approach to overcome the limitations of the Stacking Ensemble. The proposed method would result in the use of Stacking Ensemble in real-world applications. 
\section{Methodology}
\label{sec:meth}

This section provides a detailed explanation of the suggested methodology for developing an efficient explainable Ensemble Learning technique that would aid in improving performance of Supervised Learning. We first describe notations (Table ~\ref{tab:notations}) used in this section. We then show Human-Centered behavior and its relatedness to Ensemble Learning. We next map an Ensemble Learning algorithm to a Human-Centered solution followed by the implementation of Ensemble Learning utilizing Human-Centered behavior. 
\begin{center} \footnotesize
\begin{longtable}{ | m{24em} | m{11em}| } 
  \hline
   \textbf{Description} &  \textbf{Symbol} \\ 
  \hline
   Current Performance & $CP$ \\ 
  \hline
   Number of Top Scorers  & $ N $ \\ 
  \hline
   Top Scorers  &  $T (T_1,T_2.....T_N) $  \\
  \hline
    Subjects & $S (P,C,M)$  \\ 
  \hline
  Top Scorer having maximum marks & $T_{max}$  \\ 
  \hline
   Total marks of Top Scorer $T_i$ $\mid $ $T_i \in T$ & $MA_{T_{i}}$  \\ 
  \hline
     Marks of Top Scorer $T_i$ in particular subject x $\in$ S $\mid $ $T_i \in T$ & $MA_{T_{i_x}}$  \\ 
  \hline
  Feedback of any topper $T_i \in T$ & $F_{T_{i}}$  \\ 
  \hline
    Total time for taking feedback  & $T_{max}$  \\ 
  \hline
  Minimum time spent on taking feedback from one student  & $T_{feedback}$  \\ 
  \hline
   Overall positive feedback about any subject $x \in S $   & $PF_x$  \\ 
  \hline
  Subset of feedback F from some toppers $T_i \in T $  & $A$  \\ 
  \hline
    Original Dataset  & $OD$  \\ 
  \hline
    Subset of Original Dataset  & $SD(SD_1, SD_2,...SD_m)$  \\ 
 \hline
    Models  & $MO(MO_1, MO_2,...MO_k)$  \\
  \hline
    Strong base classifiers   & $B(B_1, B_2,...B_n)$ \\
       & $where$ \\
        & $0<i<m$; $0<j<k$ \\
            \hline
    Probablity Matrix  & $PM$  \\
  \hline 
     Performance Threshold  & $\lambda$  \\
  \hline 
      Extrinsic Diversity  & $D_{ext}$  \\
 \hline
    Intrinsic diversity  & $D_{int}$  \\ 
  \hline
      Clusters  & $CL(CL_1,CL_2,...,CL_c)$  \\
  \hline
     Performance of model i on dataset j  & $P(i,j)$  \\
  \hline
\caption{\label{tab:notations}Notations}
\end{longtable}
\end{center}
\subsection{Human-Centered Behaviour}
\label{sec:meth-HCB}
In this section, we relate Ensemble Learning models to scenarios that we encounter in our daily lives. Let us look at the following scenario: we need to crack the Course semester exam which has three subjects (S): Physics (P), Chemistry (C), and Math (M). We need to get feedback from the previous year's top-N scorers $T(T_1, T_2,.,T_N)$  such that our current performance (CP) can be increased to the maximum potential. Feedback from respective toppers is represented as $F (F_{T_1}, F_{T_2},...F_{T_N})$. The only constraints are i) We have a maximum of 1 hour with the top scorers $(t_{max} = 60 \: minutes)$. ii) To take feedback from any topper, it takes a minimum of 10 minutes i.e., $t_{feedback} >= 10 \: minutes$. In this time limit, we can at maximum take feedback from 6 toppers. In mathematical terms, the problem can be summarized as follows: 

\begin{center}
\begin{minipage}{0.75\linewidth}
Student’s current overall performance  = $CP$\\
Subjects, $S = (P,C,M)$ \\ 
N toppers, $T = (T_1, T_2,......T_N)$ \\
Total marks of N toppers, $MA = (MA_{T_1}, MA_{T_2},.....MA_{T_N})$\\
where $MA_{T_N} = ({MA_{T_{N_P}}}, {MA_{T_{N_C}}}, {MA_{T_{N_M}}})$ \\
Feedback from N toppers , $F = (F_{T_1}, F_{T_2}….F_{T_N})$\\ 
Overall positive feedback about any subject x $(where, x \in S)$ = $PF_x$\\
Constraints :  $t_{max}$ = 60 min, $t_{feedback}$ > = 10 min\\
A =  Subset of Feedback $F$ from some toppers  $T_i  \in T$\\
Objective : $\{ A \subseteq F | max (CP) \}$

\end{minipage}   
\end{center}

Given $CP, T, F, MA, PF_x$ and constraints ( $t_{max}$ and $t_{feedback}$), maximize the current performance $CP$ of a student by taking feedback from appropriate top scorers. In such a case, humans often react in following ways:  
 
 \begin{enumerate}[label=\roman*)]
     \item Only the top performer should be questioned about how he prepared for the exam. This can be formulated as follows: $X = \{F_{T_{max}}\}$.
     \item Solicit feedback from the top y students at the institute. This can be formulated as follows: $X = \{ F_{T_i} | i \in top \; y \; students$\}
     \item Take detailed feedback from only three students who are best in Physics ($F_{T_i} | MA_{T_{i_P}} >= MA_{T_{j_P}} \forall \: T_i ,T_j \in T$), Chemistry ($F_{T_i} | MA_{T_{i_C}} >= MA_{T_{j_C}} \forall \: T_i ,T_j \in T$), and Math ($F_{T_i} | MA_{T_{i_M}} >= MA_{T_{j_M}} \forall \: T_i ,T_j \in T$) irrespective of their rank. This can be formulated as follows: $X = \{(MA_{T_{i_P}} >= MA_{T_{j_P}} \forall \: T_i ,T_j \in T$), ($MA_{T_{i_C}} >= MA_{T_{j_C}} \forall \: T_i ,T_j \in T$), ($MA_{T_{i_M}} >= MA_{T_{j_M}} \forall \: T_i ,T_j \in T$\}
 \end{enumerate}

Let us evaluate all three cases. 

\begin{figure}[!h]
  \centering
  \includegraphics{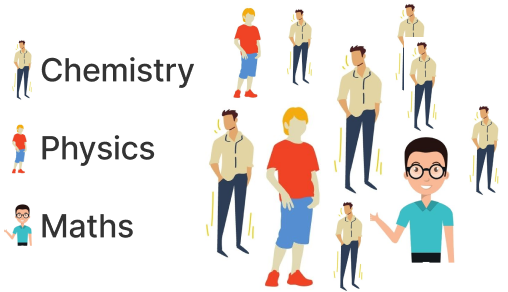}
  \caption{Case II: Ask Top-N Toppers}
  \label{fig:Case_ii- Ask top n}
\end{figure}

Case I) $X = \{F_{T_{max}} \}$ \\
Problem with this case is : $MA_{T_{max}} >= MA_{T_i} \forall i \in T, \exists T_j \in T | MA_{T_{j_C}} > MA_{T_{max_C}}$ i.e a topper may be exceptional in Physics and Math but only marginally so in Chemistry. In such a case $PF_P = PF_M > PF_C$.  But there might be another student who might be exceptional in Chemistry but average/below average in other subjects. In such a case, getting feedback on Chemistry from a student who excels at it would be preferable to getting it from a top-performing student. As a result, relying solely on one student may not be a wise choice. It may result in the formation of a bias towards a particular subject.
 \begin{figure}[!h]
  \centering
  \includegraphics{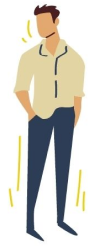}
  \caption{Case I: Ask Only the Topper}
  \label{fig:Case i: Ask only the topper}
\end{figure}
Case II) $X = \{ F_{T_i} | i \in top \; y \; students$\} \\
Such a scenario may appear to be efficient, but it is not. Suppose, the following was the difficulty of the paper which toppers gave - $C << P < M$. In such a case, $MA_{T_{i_C}} >>> MA_{T_{i_M}} \forall T_i \in T$.  Most toppers (e.g., 7/10) are likely to find Chemistry easy and interesting, while Math is likely to be difficult. In such a case, $PF_C >> PF_P > PF_M$. Hence, feedback taken from only toppers might lead to creating a bias towards a particular subject (Because the paper is the same for all, most of the top performers think similarly and have similar weaknesses and strengths). But next year, if $C >> P > M$, then it might become a problem as students might have studied Chemistry very well and not focused much on Math (because of biased feedback which we had received). As a result, this approach is also inefficient. Also, we may be perplexed as to how many toppers to solicit feedback from. This may necessitate devoting less time to each topper (e.g., 10 minutes) in order to obtain the most feedback. Figure \ref{fig:Case_ii- Ask top n} explains the same thing visually.

\begin{figure}[!h]
  \centering
  \includegraphics[width=4.cm]{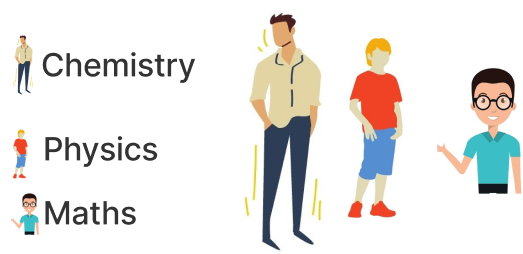}
  \caption{Case III: Take Feedback Only from Top Scorers in Each Subject}
  \label{fig:case_iii - Ask best.png}
\end{figure}

Case III) $X = \{(MA_{T_{i_P}} >= MA_{T_{j_P}} \forall \: T_i ,T_j \in T$), ($MA_{T_{i_C}} >= MA_{T_{j_C}} \forall \: T_i ,T_j \in T$), ($MA_{T_{i_M}} >= MA_{T_{j_M}} \forall \: T_i ,T_j \in T$\} \\
This turns out to be the most logical and efficient method, both in terms of efficiency and time. In this case, $PF_C = PF_P = PF_M$, unlike the above-mentioned cases ($PF_P = PF_M > PF_C$ in case i and $PF_C >> PF_P > PF_M$  in case ii ) where feedback is biased towards a particular subject. Unbiased feedback is possible because all three students have different areas of expertise and think in different ways. So, even if the exam's difficulty level changes next year, it won't make much of a difference in our performance because we received unbiased feedback. We can devote more time to each topper (twenty minutes per top scorer) and get quality feedback because we know exactly how many top scorers to take feedback from (three). The same idea is illustrated in Figure \ref{fig:case_iii - Ask best.png}.

According to the preceding example, rather than going for a large number of people who have done well (which may require time, and resources and may also create bias at times), it is better to limit to fewer people whose performances are adequate and have diverse opinions or more diversity. So, in this case, diversity can be defined as soliciting feedback only from students who think differently in order to achieve a common goal, i.e. to score the highest possible marks.Because there are only three subjects in the exam, we can say that students who excel in different subjects think differently. As a result, taking feedback only from top scorers in individual subjects will provide maximum diversity while taking the least amount of feedback.

\subsection{Mapping Ensembles to Human-Centered Behaviour}
\label{sec:meth-Map-Ensemble-HCB}

The aforementioned problem is very similar to Ensemble Learning. Ensemble Learning combines multiple strong base classifiers to make correct predictions as discussed in the above Feedback example. Merging models solely on their performance may lead to bias if many high-performing models make similar predictions, as in case(II) of Section \ref{sec:meth-HCB}. Choosing a large number of base learners is also impractical due to time and resource constraints, as in case(iii) of Section \ref{sec:meth-HCB}. As a result, similar to case(III) of Section \ref{sec:meth-HCB}, we must select the bare minimum of base models that behave differently. This will make Ensemble Learning more efficient, explainable, and feasible to implement in practice. As in case(III) of Section \ref{sec:meth-HCB}, we can say that in Ensemble Learning, maximum performance can be achieved efficiently by using a minimal number of models that are more diverse. Diversity, as defined under Section \ref{sec:meth-HCB}, can be defined as taking only those base models that differ significantly in how they make predictions in order to achieve the desired goal, i.e., maximum performance. However, before using this equation, two major issues must be addressed: 
\begin{enumerate}[I]
    \item We already knew the Toppers $T(T_1, T_2…T_n)$ who was in the problem discussed in Section \ref{sec:meth-HCB}. Here, we need to find all of the top-performing base classifiers $B(B_1, B_2,...B_n)$, which can provide us with as much diversity as possible.
    \item In the real-life example, diversity was simple to define because we already knew there were three subjects $S(P, C, M)$. To achieve the desired diversity, we can consider toppers in each subject. In this case, however, we must first determine how models can be classified into groups G. We can select high-performing models from each group G after we divide models into C groups.
\end{enumerate}
We define two terms: Extrinsic diversity ($D_{ext}$) and Intrinsic diversity ($D_{int}$) to address these two difficulties. After we have resolved this problem, we can use logic similar to the example in Section \ref{sec:meth-HCB}.

\subsubsection{Extrinsic Diversity ($D_{ext}$)}
\label{sec:Dext}
Extrinsic Diversity ( $D_{ext}$ ) refers to identifying a set of strong base classifiers $B(B_1, \\B_2…B_n)$ that meet a predefined performance threshold $\lambda$. It is called $D_{ext}$ because we choose all of the possible base classifiers whose performance is greater than $\lambda$ which can provide diversity by looking at external factors. External factors refer to claiming diversity by investigating the mathematics of various Machine and Deep Learning algorithms, as well as feature selection techniques. We call it Extrinsic diversity because we make decisions based on our understanding of math rather than delving into the details of how a particular model makes predictions for given instances of the dataset. We can relate it to finding toppers $T(T_1, T_2,...T_n)$ and their feedback $F(F_1, F_2,...F_n)$ of the above example. $D_{ext}$ helps us to solve problem I, which we discussed in Section \ref{sec:meth-Map-Ensemble-HCB}.

$D_{ext}$ is achieved by first identifying subsets $SD (SD_1, SD_2,...SD_m)$ of the original dataset $OD$ using a variety of feature reduction and extraction techniques, and then applying various algorithms to these subsets of datasets. The idea behind this is to get as many diverse base models as possible. The detailed steps for finding $D_{ext}$ are as follows:

\begin{enumerate}[label=\Alph*.]
    \item Creating multiple subsets from the original dataset \\ \\
        To achieve maximum diversity, we first use various feature extraction and reduction techniques to find $m$ subsets of the original dataset $SD(SD_1, SD_2,...SD_m)$ with different features. The goal is to have as much variety as possible in the datasets. We assume that running models on different subsets of the dataset SD will give us more variety in terms of results. Principal component analysis (PCA), Recursive Feature Elimination (RFE), Logistic Regression, and Feature Selection using Distinct Basic Analysis Strategies are the techniques we employed to produce various subsets with varied features.
    \item  Defining Strong Base Classifiers \\ \\
    We obtained the various datasets that we required from above. We also require a diverse set of models that predict instances in different ways. We then apply all of the models, from simple Naive Bayes to Neural Networks, to all subsets of original datasets that we obtained. To achieve diverse predictions, we used the following models: Naive Bayes, Logistic Regression, SVM, KNN, Decision Tree, Random Forest, XGBoost, and Neural Networks. Grid Search is used to determine the best parameters for all algorithms. All these models are stored in a list called ‘Models’. We form $k$ such strong models $MO(MO_1,MO_2,...MO_k)$.\\


    \item Algorithm
    
    After we have formed various subsets of the dataset and defined models, we then use the $D_{ext}$ algorithm to find all (model, dataset) pairs whose performance exceeds a certain threshold. The $D_{ext}$ algorithm is explained in detail below.

\begin{algorithm}[!h]
\caption{$D_{int}$ Algorithm}\label{alg:Human-Centered Ensemble Algorithm}
\begin{algorithmic}[1]
\State $B \gets  []$
\State $B \gets  Empty \;\; DataFrame$
\State $Models \gets MO(MO_1, MO_2,...MO_k)$    
\State $Subsets \gets SD(SD_1, SD_2,...SD_m)$  
\State $\lambda \gets$ Threshold value

\For {$j=1,m,\ldots$}
    \For {$i=1,k,\ldots$}
        \State Apply algorithm i on the dataset j to compute Performance $P(i,j)$
        \If{$P(i,j) > \lambda$ }
            \State $B.append((j,i))$
            \State $PM.addRow(prediction_prob(j,i,instances,classes))$  
        \EndIf
    \EndFor
\EndFor
\State $CL = (CL_1,CL_2,...CL_n)$
\State $I = n+1 $

\end{algorithmic}
\end{algorithm}    
    
    Line 1 creates an empty set $B$, which holds pairs of strong models and subsets. Line 2 initializes the empty data frame for the Probability matrix $PM$. $PM$ is used to save class predictions for each instance of the selected model. We then create a list containing all the strong models $MO (MO_1, MO_2,...MO_k)$ with the best hyper-parameters found by applying grid search. Line 4 defines 'subset' as a list containing a subset of the original dataset $SD(SD_1, SD_2,...SD_m)$ obtained using the various feature selection and extraction techniques. Line 5 defines a minimum threshold performance $\lambda$. The for loop of lines 6 - 14 iterates through every subsets $j$ in the Subsets. For each subset j, the for loop of lines 7 - 13 iterates through every model $i$ in the Models. Line 8 computes the performance of model $i$ for the current subset $j$ ( P(i,j) ).  If P(i,j) > $\lambda$ then we execute lines 10 and 11. Line 10 appends a pair of subset j and model I from set $B$. On line 11, we append predicted class probabilities on a training set for model $i$ to dataset $j$ in $PM$. The ultimate goal of $D_{ext}$ is to construct a Probability matrix $PM$ and identify a set of $n$ strong (Models,Subset) pairs $B (B_1, B_2…B_i...B_n)$.


    \item More about B and PM \\ \\
    The following section delves more into what B and PM are.
        \begin{enumerate}[label=\roman*.]
            \item Strong Base Classifier ($B$) 
    
    Strong Base Classifiers ($B$) is a list, which contains a set of Model and Subset as its element. Consider the example shown in Figure \ref{Strong Base Classifiers (B)}. Using the above mentioned algorithm, we first formed $m$ subsets $SD(SD_1, SD_2,...SD_m)$. We then defined $k$ base models $MO(MO_1, MO_2,...MO_k)$. We then go through every possible combination of subsets and models to find high-performing pairs of $(MO, SD)$ with performance greater than $\lambda$. These high-performing pairs can be found in $B$.
        \begin{figure}[!h]
              \centering              \includegraphics[width=14.2cm]{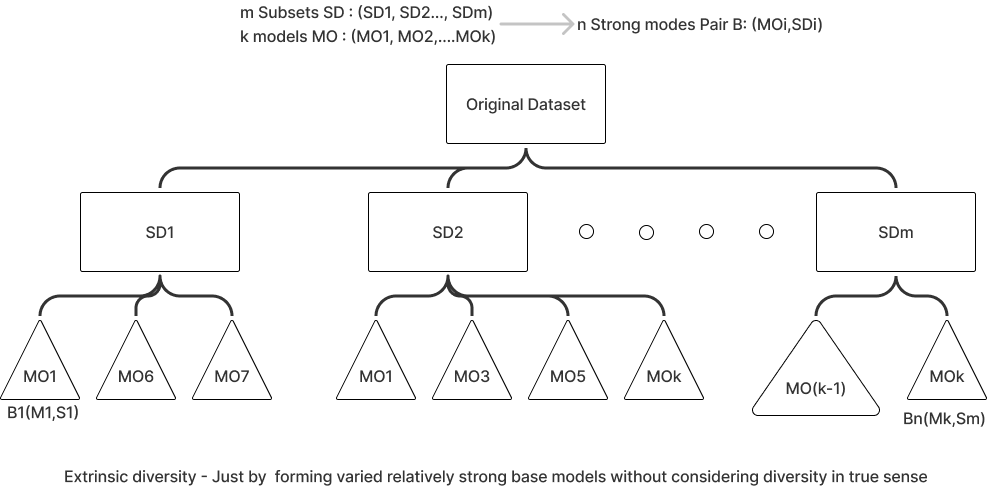}
              \caption{Strong Base Classifiers (B)}
              \label{Strong Base Classifiers (B)}
            \end{figure}
    Figure \ref{Strong Base Classifiers (B)} shows all the models for a subset that crosses threshold $\lambda$. From Figure \ref{Strong Base Classifiers (B)}, we can say that $B$ = [
    $\{MO_1,SD_1\}, \{MO_6,SD_1\}, \{MO_7,SD_1\}, \{MO_1,SD_2\}$, $\{MO_3,  SD_2\}, 
    \{MO_K,SD_2\},\hdots,\{MO_1,SD_m\}, \{MO_k-1,SD_m\}]$. Each such set of $(MO, SD)$ is called a strong model B. Hence $B = (B_1,B_2,...B_n) $, where $B_1 = \{MO_1,SD_1\}$, $B_2 = \{MO_6,SD_1\}$,... $B_n = \{MO_k,SD_m\}$. 

            \item Probability Matrix (PM) 
            
    The Probability Matrix ($PM$) is (n x t ) matrix where n = number of Base Classifiers $B$ obtained from $D_{ext}$ i.e $B(B_1,B_2,...B_n)$, t = i * c , c = number of classes, i = number of instances. $PM$ gives an idea of how predictions of different classifiers in $B$ differ. Every row of $PM$ corresponds to an element in $B$. Columns represent the probability of instances belonging to a specific class. For example, if there are 'c' classes, the first 'c' columns represent the likelihood of a first instance belonging to each class for each model in B. We know that there are ‘i’ instances, hence dimensionality would be [n X t]. (Figure \ref{Probability Matrix Structure})
    
    \begin{figure}[!h]
        \centering
        \includegraphics[width=8cm]{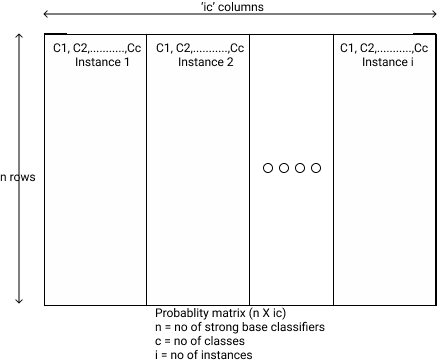}
        \caption{Probability Matrix Structure}
        \label{Probability Matrix Structure}
    \end{figure}

Extrinsic Diversity may not provide us with truly diverse models because while they differ in mathematical aspects, they may be similar in terms of making predictions. The end goal of $D_{ext}$ is to find a set of strong models, subset pairs $B(B_1, B_2…B_n)$ and build a Probability matrix $PM$. We can derive a minimum subset of $B$ from this $PM$ that will provide maximum diversity while also improving performance efficiently. It can be achieved using Intrinsic Diversity $D_{int}$.
    
        \end{enumerate}
    \end{enumerate}
\subsubsection{Intrinsic Diversity ($D_{int}$)}
\label{sec:Dint}
Intrinsic factors refer to delving deeper into how each model makes predictions and then determines the smallest subset of $B$ that provides the maximum diversity using $PM$. This problem is analogous to first identifying all possible groups (such as $S(P, C, M)$ in Section \ref{sec:meth-HCB}) and then selecting only high-performing diverse models, as in case $III$ of Section \ref{sec:meth-HCB}. $D_{ext}$ assists us in resolving problem $II$, which was discussed in Section \ref{sec:meth-Map-Ensemble-HCB}.

\begin{enumerate}[label=\Alph*.]
            \item Significance of $PM$
            
            \begin{figure}[!h]
              \centering
              \includegraphics[width=12cm]{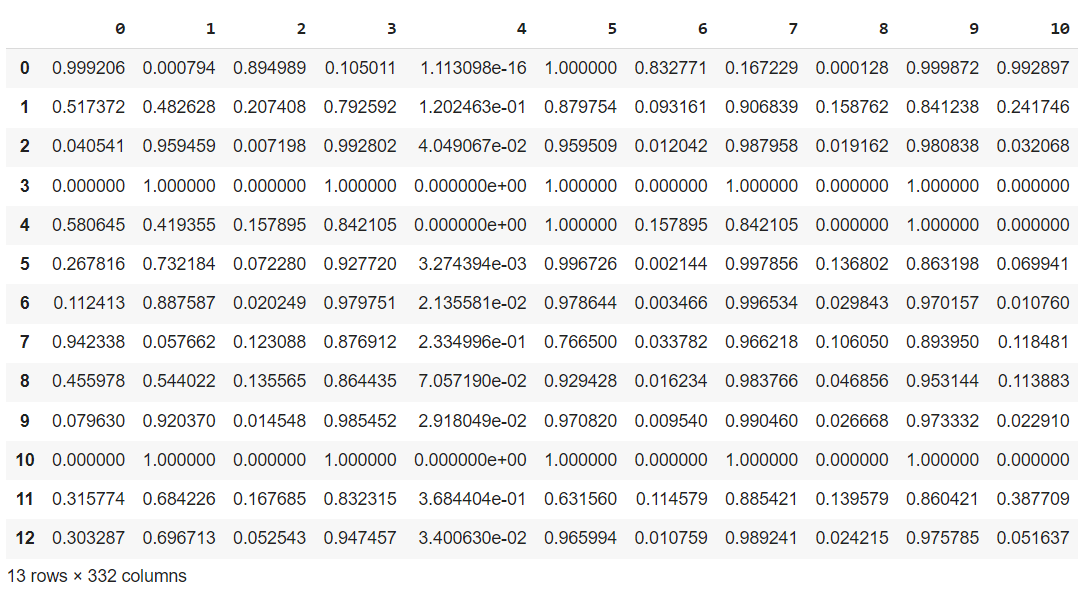}
              \caption{Probability Matrix}
              \label{Probability matrix}
            \end{figure}        
            This section explains the logic behind using $PM$ to divide strong classifiers $B$ into groups. Consider the small PM in Figure \ref{Probability matrix} as an example. In the above Figure, m = 13 = no of base models; c = 2 (as it is 2 class classification problem); n = 166 (no of instances). The first two columns show how all thirteen base modes classified the first instance of train data. For explanation purposes, consider rows 0, 1, and 7 and the first two columns. It represents how models 0,1 and 7 classified the first instance of the dataset. As the probability of class 0 is greater than 0.5 in all of them, we can conclude that all these models classified it as class 0. However, models 0 and 7 had a probability of 0.9+ when classifying an instance as class 0, whereas model 1 had a probability of only 0.52. While all three models correctly classified instances, model 1 differs significantly from the other two. To categorize models, we used the probability concept discussed above. We simply need to apply this logic to each model and each instance. To accomplish this task, we use the concept of hierarchical clustering on PM and then divide all strong base classifiers int models in $c$ groups. The optimum value of $c$ is determined using dendrogram analysis. Hence, we can now divide all the rows of $PM$ (Strong classifiers $B$) into groups and select only the best-performing sets in $B$ from each group to form the final Ensembles.      
            \item Algorithm
            
            The first step of the $D_{int}$ approach is to apply a hierarchical clustering algorithm on $PM$ to create a dendrogram. Following that, an appropriate number of clusters are formed based on dendrogram analysis.         
\begin{algorithm}[!h]
\caption{$D_{int}$ Algorithm}\label{alg:Human-Centered Ensemble Algorithm}
\begin{algorithmic}[1]
\State $PM \gets $ Probability matrix obtained from $D_{ext}$
\State $dist(cl_1,cl_2) \gets $ Euclidean Distance between 2 points
\State $CL \gets []$
\For {$i=1,n,\ldots$}
    \State $CL_i = \{B_i\}$
\EndFor
\State $CL = (CL_1,CL_2,...CL_n)$
\State $I = n+1 $

\While{$Size(CL) > 1$}
    \State $(cl_{min1},cl_{min2})$ = minimum $dist(cl_i,cl_j) \;\;\forall \;cl_i,cl_j \in CL$
    \State Remove $cl_{min1}$ and $cl_{min2}$ from $CL$
    \State Add $(clmin1,clmin2)$ in $CL$
    \State $I = I + 1$
\EndWhile

\State Analyze dendrogram and define threshold
\State Divide into k clusters by threshold

\end{algorithmic}
\end{algorithm}
            
            Lines 1, 2, and 3 define the Probability Matrix, Euclidean Distance, and Clusters respectively. Lines 4-6 of the for loop initialize all the rows of $PM$ (all $n$ strong base models $B$) as individual clusters $CL$.So, initially, we get $n$ clusters as shown in line 7. Lines 9-14 contain a while loop that runs until we merge all of the clusters in $CL$ to form a single cluster and obtain a dendrogram. Line 10 computes the distance between each cluster and finds clusters with the shortest distance, i.e. models that think similarly based on classification probabilities. Lines 11 and 12 remove the individual clusters and join them to form a new cluster.

            We then plot a dendrogram of the same after we merge all the clusters and form a single cluster, i.e., after the while loop ends. We try to analyze the pattern in the dendrogram and define a threshold value for splitting the dendrogram that will give us $c$ groups into which we can classify all models. If no definite pattern is found, this threshold is sometimes taken using the try and test technique. The dendrogram formed using the above-mentioned algorithm for the $PM$ shown in Fig \ref{Probability matrix} is shown below in Fig \ref{Dendrogram}.         
            \begin{figure}[!h]
              \centering
              \includegraphics[width=10cm]{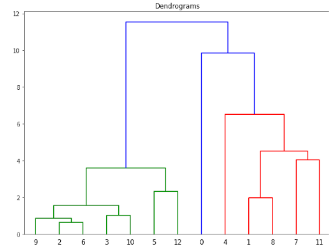}
              \caption{Dendrogram}
              \label{Dendrogram}
            \end{figure}
            This gives a fair idea of how many clusters (groups) to form. A threshold value is defined next where we decide to split the clusters. In this dendrogram, we chose a threshold of five because it produced the best results.

            \begin{figure}[!h]
              \centering
              \includegraphics[width=10cm]{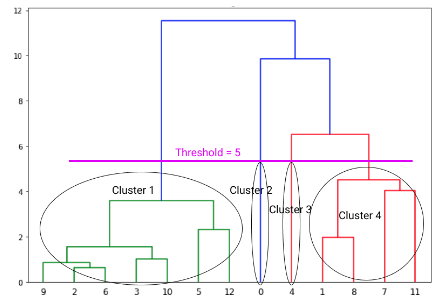}
              \caption{Dendrogram Split by Threshold}
              \label{Splitting Dendrogram by setting threshold}
            \end{figure}
            When the threshold was set to five, four clusters (c = 4) as shown in Figure \ref{Splitting Dendrogram by setting threshold} were formed. By having a look at the dendrogram from a distance, we can get an idea about how many clusters are required to be formed. This becomes an important consideration when determining the value of the threshold.

            The two problems discussed in Section \ref{sec:meth-Map-Ensemble-HCB}, which prevented us from using the Human-Centered logic we used in Section \ref{sec:meth-HCB} have been solved in Section \ref{sec:Dext} and \ref{sec:Dint}. We now form the human-centered ensemble algorithm using the similar logic described in Section \ref{sec:meth-HCB}.

\end{enumerate}
\subsubsection{Human-Centered Ensemble Algorithm}

First, we looked at how $D_{ext}$ may be utilized to identify robust base classifiers. Then, all of these strong base classifiers were divided into groups using the $D_{int}$ algorithm. In this section, we will examine how we may build on these concepts to create a stacking ensemble utilizing a proposed Human-Centered Ensemble Algorithm that uses fewer base models.
\begin{algorithm}
\caption{Human-Centered Ensemble Algorithm}\label{alg:Human-Centered Ensemble Algorithm}
\begin{algorithmic}[!h]
\State $Perform \; \; basic \; \;  preprocessing$
\State $L1 \gets []$
\State Find $D_{ext}$ 
\State Find $D_{int}$ 
\State $Final\_Base\_Models \gets []$

\For {$i=1,c,\ldots$}
    \State append($Final\_Base\_Models$ , $max\_performing\_model(i)$)
\EndFor

\For {$i=1,len(Final\_Base\_models) ,\ldots$}
    \State add($L1$, $pred(i)$) 
\EndFor

\State Apply the highest performing algorithm on the newly formed dataset to make predictions

\end{algorithmic}
\end{algorithm}

 Algorithm \ref{alg:Human-Centered Ensemble Algorithm} represents the proposed novel Human-Centered algorithm. Line 1 executes the fundamental cleaning operations. Line 2 declares the empty data frame L1. Once we get a cleaned dataset, Line 3 finds the Extrinsic diversity $D_{ext}$ discussed in Section \ref{sec:Dext}. The output of $D_{ext}$ is $PM$ and $B$. Line 4 exemplifies the Intrinsic diversity discussed in Section \ref{sec:expevl}. Using the concept of hierarchical clustering, $D_{int}$ assists us in grouping all of the similar models in B into 'c' groups. Line 5 declares a list ‘$Final\_Base\_Models$’, where we store the final minimal base models selected for Ensemble Learning. The for loop of line 6 adds all the high-performing models of each group into $Final\_Base\_Models$, which is similar to case iii of Section \ref{sec:meth-HCB}. The for loop of line 9 appends the probability predictions of all strong models (present in $Final\_Base\_Models$) for all the instances of the L1 dataset. An algorithm that performs best is then used in the level 1 dataset in order to obtain the final prediction.

\begin{figure}[!h]
  \centering
  \includegraphics[width=12cm]{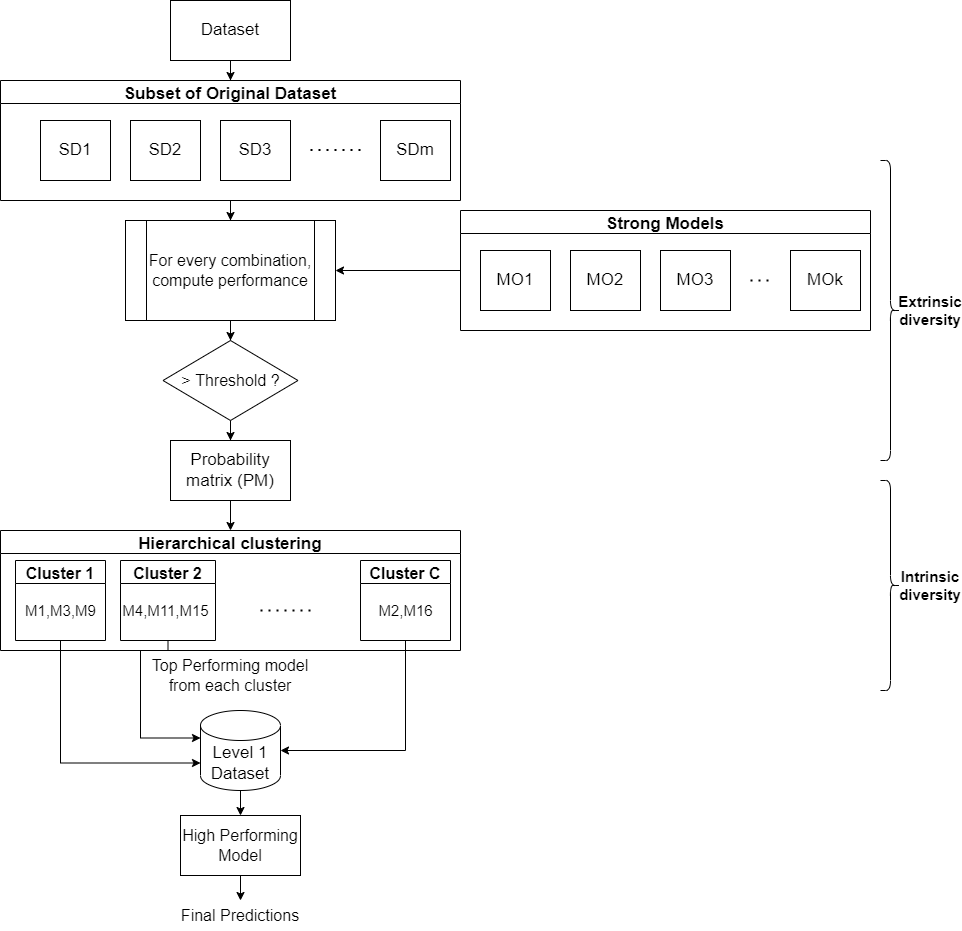}
  \caption{Flowchart of Proposed Model}
  \label{Proposed model flowchart}
\end{figure}

Figure \ref{Proposed model flowchart} depicts the above-mentioned algorithm. We begin by evaluating the dataset and performing basic processing on it. Then, based on the predicted satisfactory performance for the given problem, a threshold is set. We then compare the performance of all the chosen models $MO(MO_1,MO_2,...MO_k)$ on obtained subsets of dataset $SD(SD_1,SD_2,...SD_m)$  with the specified threshold to find all the strong base classifiers B.  We add $(MO_i, SD_j)$ to strong base classifier B if performance for a given model $MO_i$ for a subset $SD_j$, i.e., P(i,j) is better than threshold. In addition, we add predictions for each class of each instance of all elements in B in PM at the same time. After that, we separate strong base classifiers into 'c' groups using hierarchical clustering on the $PM$. The best models from each group are then used as base classifiers in Ensemble Learning. By adding predictions from specified base models, we create a level one dataset. Finally, the highest performing algorithm on the level one dataset is used to make the prediction.

Therefore, in this study, we sought to comprehend how applying human-centered concepts might enhance the performance of Supervised Learning. We discovered that while Stacking Ensembles learning generally enhances performance, it has a few limitations that prevent it from being extensively adopted. Therefore, by studying human-centered behavior and using a similar analogy to Ensemble Learning, we tried to make Stacking Ensemble Learning more comprehensible while simultaneously decreasing computing costs. Results demonstrated that the suggested approach performs better than the suggested models and is also more practical when taking into account real-life scenarios.

\section{Experimental Evaluation}
\label{sec:expevl}
This section begins with an overview of the datasets that were utilized to test the proposed method. The performance of the suggested model is compared against the conventional approaches of Ensemble Learning and to that of recent research papers' state-of-the-art results, which are then displayed in Inter-model results. Finally, intra-model findings are shown to depict the performance of the proposed approach, thus illustrating its importance.

\subsection{Dataset Description}
Experiments were carried out on nine datasets (which are mentioned below) to assess the proposed method's performance. Proof-of-concept was tested using the first three datasets. This lets us figure out whether what we are proposing is working or not. The same experiment was then performed on the remaining six datasets and the results of these were compared to benchmark findings published in recent research papers for the remaining six datasets. 
 \begin{enumerate}[label=\roman*.]
     \item Breast Cancer Wisconsin dataset: 
The goal was to classify tumors as either malignant (cancerous) or benign (non-cancerous). There were 570 samples in the dataset. There were 63\% benign tumors and 37\% malignant tumors.
    \item Sonar dataset: 
The objective was to determine whether the given sample was Rock (R) or Mine (M). There were 208 samples in the dataset. Sixty-six percent were Rock and forty-four percent were Mine.
    \item Wine dataset:
The aim was to classify the wine's quality (0,1,2). There were 535 samples in the dataset. The data was nearly balanced and cleansed.
    \item Diabetic Retinopathy Data Set:
This dataset comprises features extracted from the Messidor picture collection that can be used to determine whether or not an image contains diabetic retinopathy indications. There are 1151 instances and 20 attributes in the collection. The ultimate goal was to determine whether or not a given instance had indications of diabetic retinopathy. With a ratio of 52:48, the dataset was nearly balanced.
    \item NSL KDD (5 - Class) Dataset:
The well-known NSL-KDD data set is useful for evaluating various intrusion detection techniques. The goal is to determine whether the provided instance is normal, dos attack, probe attack, privilege attack, or access attack. The distribution of different forms of data is unbalanced, with Normal instances accounting for the majority of overall data.
    \item NSL KDD (2 - Class) dataset: 
The well-known NSL-KDD data set is useful for evaluating various intrusion detection techniques. The goal here was to determine whether a particular instance is normal or suspicious.
    \item Covid-19 dataset: 
This information was gathered from 5644 patients who were admitted to the Albert Einstein Israelita Hospital. 559 patients out of 5644 tested positive for covid. The goal here was to determine whether or not a certain instance is affected by covid.
    \item Dry beans dataset:
This dataset is made up of 15 features extracted from soya bean images. There are 13,611 grains in this dataset, which are divided into seven categories. The purpose was to appropriately classify grains into the appropriate category.
    \item HTRU2 Dataset:
The HTRU2 data set contains a selection of pulsar candidates discovered during the High Time Resolution Universe Survey. Pulsars are a type of Neutron star that emits radio waves that can be detected on Earth. There are 16,259 erroneous cases generated by RFI/noise and 1,639 actual pulsar examples in the data set supplied here. Human annotators have double-checked each of these cases. Our goal is to determine whether or not a given event is a genuine pulsar.
 \end{enumerate}
 \subsection{Inter-model Results}
In this section, we compare the proposed methods' experimental findings to extensively used and verified approaches. First, we compare the performance of our models to that of the most extensively used Stacking Ensemble method. Finally, we compare the performance of our model with the state-of-the-art model’s performance proposed in recent research papers.
\begin{enumerate}[label=\Alph*.]
\item Comparing Performance against Conventional Approach\\ \\
Only high-performing models are frequently used as base models in Ensemble Learning. In our situation, we chose models that were diverse rather than only adopting high-performing models. We also compared our strategy to the traditional method of selecting top performing $n$ models. The number of base models obtained after clustering was chosen as the value of $n$. 
\begin{table}
\scriptsize
\centering
\caption{Performance Comparison of Proposed Approach with Conventional Approaches}
\label{tab:Conventional approach}
\resizebox{\textwidth}{!}{  
\begin{tabular}{|l|c|c|} 
\hline
\textbf{Dataset} & \textbf{\begin{tabular}[c]{@{}c@{}}Conventional Approach \\ Perf. (\%)\end{tabular}} & \textbf{\begin{tabular}[c]{@{}c@{}}Proposed Approach \\ Perf. (\%)\end{tabular}} \\ 
\hline
Breast cancer (1995)      & 97.9  & 100   \\ 
\hline
Sonar                     & 85.7  & 90.5  \\ 
\hline
Wine (1991)               & 100   & 100   \\ 
\hline
Diabetic Retinopathy (2014) & 77.9  & 82.3  \\ 
\hline
NSL KDD (5-class, 1999)  & 84.3  & 90.1  \\ 
\hline
NSL KDD (2-class, 1999)  & 83.2  & 85.2  \\ 
\hline
Covid-19 (2020)           & 99.91 & 100   \\ 
\hline
Soya bean (2020)           & 93.09 & 93.46 \\ 
\hline
HTRU2 (2017)              & 98.53 & 98.53 \\ 
\hline
\end{tabular}
}
\end{table}
Table \ref{tab:Conventional approach} shows the performance of the conventional strategy versus our approach. From Table \ref{tab:Conventional approach}, we conclude that the proposed technique outperformed the standard approach in terms of results.
\item Performance Comparison against State-of-the-art Approaches\\ \\
In this section, we present the performance improvement obtained using our proposed approach against the state-of-the-art approaches on six datasets. 
Figure~\ref{Fig: Base model Vs Proposed model performance} illustrates the performance of the state-of-the-art method against our proposed method.

The performance of the state-of-the-art method versus our approach is shown in Table \ref{tab:state-of-art approach}. 
\begin{table}[!ht]
\centering
\caption{Performance Comparison of Proposed Approach with State-of-the-art Approaches}
\label{tab:state-of-art approach}
\resizebox{\textwidth}{!}{  
\begin{tabular}{|l|l|c|c|} 
\hline
\textbf{Dataset} & \textbf{Methods} & \textbf{\begin{tabular}[c]{@{}c@{}}State-of-the-art \\ Perf. (\%)\end{tabular}} & \textbf{\begin{tabular}[c]{@{}c@{}}Proposed model \\ Perf. (\%)\end{tabular}} \\ 
\hline
HTRU2 (2017)              & RTB-VC, Rustam et al. (2020)       & 98.2  & 98.5  \\ 
\hline
Soya bean (2020)           & SVM, Koklu et al. (2020)           & 93.13 & 93.46 \\ 
\hline
Diabetic Retinopathy (2014) & RF+DT+Adaboost+KNN+LR, Reddy et al. (2020) & 81.9  & 82.3  \\ 
\hline
Covid-19 (2020)           & Extra tree+RF+LR, AlJame et al. (2020)  & 99.94 & 100   \\ 
\hline
NSL KDD (5-class, 1999)  & Adaptive Machine learning model, Xianwei & 85.2  & 90.1  \\ 
\hline
NSL KDD (2-class, 1999)  & SAE\_SVM, Al-Qatf et al. (2018)    & 84.96 & 85.2  \\ 
\hline
\end{tabular}
}
\end{table}
Our proposed model achieves an improvement of 0.30\% on HTRU Dataset'17, 0.33\% on Soya bean Dataset'20, 0.4\% on Diabetic Retinopathy dataset’14, 0.60\% on Covid-19 Dataset’20, 4.9\% on NSL KDD Dataset’99 (5-
class) and 0.24\% on NSL KDD Dataset’99 (2-class) respectively.
\end{enumerate}

 \begin{figure}[!ht]
  \centering  \includegraphics[width=0.85\textwidth]{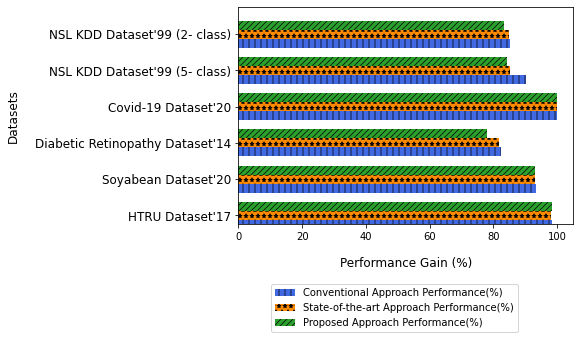}
  \caption{Performance Comparison of Conventional Approaches, State-of-the-art Approaches and Proposed Approach}
  \label{Res: SOTA Vs Con Vs Prop}
\end{figure}
Figure \ref{Res: SOTA Vs Con Vs Prop} shows a performance comparison of the conventional approach, state-of-the-art approach, and proposed approach which was discussed above. This confirms that the proposed approach outperforms both traditional and state-of-the-art approaches.
\subsection{Intra-model results}
This section compares intra-model findings on two crucial parameters: time and performance, to show the value that the proposed approach may provide in real-world applications. This section also gives a concise summary of how the proposed method might achieve the needed balance of time, performance, and interpretability.
\begin{enumerate}[label=\Alph*.]
\item Performance Evaluation \\ \\
The hypothesis we intended to explore was if merging just a few minimum base models could improve the performance of the best base model. We utilized the approach indicated in Section \ref{sec:meth} on the nine datasets provided to test this hypothesis.
\begin{figure}[!h]
  \centering
\includegraphics[width=0.85\textwidth]{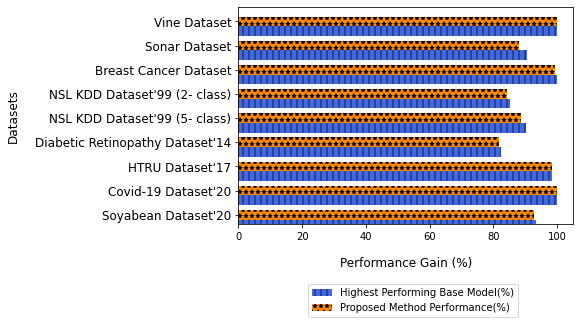}
  \caption{Performance Comparison of the Strongest Base Model with the Proposed Model}
  \label{Fig: Base model Vs Proposed model performance}
\end{figure}
\begin{table} \scriptsize
\centering
\caption{Performance Comparison of Proposed Model with Strongest Base Model}
\label{tab:Base Model vs Proposed Model Performance}
\resizebox{\textwidth}{!}{  
\begin{tabular}{|l|c|c|c|} 
\hline
\textbf{Dataset} & \textbf{\begin{tabular}[c]{@{}c@{}}Highest base \\ model perf. (\%)\end{tabular}} & \textbf{\begin{tabular}[c]{@{}c@{}}Base models \\ taken\end{tabular}} & \textbf{\begin{tabular}[c]{@{}c@{}}Proposed model \\ perf. (\%)\end{tabular}} \\ 
\hline
Soya bean (2020)           & 92.76 & 3 & 93.46 \\ 
\hline
Covid-19 (2020)           & 99.8  & 4 & 100   \\ 
\hline
HTRU2 (2017)              & 98.3  & 3 & 98.5  \\ 
\hline
Diabetic Retinopathy (2014) & 81.8  & 3 & 82.3  \\ 
\hline
NSL KDD (5-class, 1999)  & 88.5  & 4 & 90.1  \\ 
\hline
NSL KDD (2-class, 1999)  & 84.4  & 4 & 85.2  \\ 
\hline
Breast cancer (1995)      & 99.3  & 3 & 100   \\ 
\hline
Sonar                     & 88.1  & 3 & 90.5  \\ 
\hline
Wine                      & 100   & 2 & 100   \\
\hline
\end{tabular}
}
\end{table}
Table \ref{tab:Base Model vs Proposed Model Performance} demonstrates that our hypothesis was correct and that our model behaved as expected. After combining up to four base models in each dataset, the performance of the highest performing base model was either unchanged or improved.
Figure \ref{Fig: Base model Vs Proposed model performance} makes the same point more comprehensible.
\item Time-Performance Trade-off \\ \\
The next point we tried to answer was whether the improvement over the base model was significant and if it could be implemented in practice. We conducted experiments to determine a time-performance tradeoff to remedy this.\\ 
The results are shown in Table~\ref{tab:time-performance tradeoff}.
\begin{table}[!ht]
\centering
\caption{Time Performance Tradeoff - Single Strong Base Model vs Ensemble Model}
\label{tab:time-performance tradeoff}
\resizebox{\textwidth}{!}{  
\begin{tabular}{|l|c|c|c|c|c|} 
\hline
\textbf{Dataset} & \textbf{\begin{tabular}[c]{@{}c@{}}Highest base \\ model time (s)\end{tabular}} & \textbf{\begin{tabular}[c]{@{}c@{}}Highest base \\ model perf. (\%)\end{tabular}} & \textbf{\begin{tabular}[c]{@{}c@{}}Ensemble \\ time (s)\end{tabular}} & \textbf{\begin{tabular}[c]{@{}c@{}}Ensemble \\ perf. (\%)\end{tabular}} & \textbf{\begin{tabular}[c]{@{}c@{}}Perf. (\%) - \\ time (s) tradeoff\end{tabular}} \\ 
\hline
Soya bean (2020)          & 42.7  & 92.7 & 53.3 & 93.46 & 0.7 $\rightarrow$ 10.6                                                                         \\ 
\hline
Covid-19 (2020)          & 9.81  & 99.8 & 10.96 & 100 & 0.2 $\rightarrow$ 1.15                                                                         \\ 
\hline
HTRU2 (2017)             & 6.24  & 98.3 & 7.13 & 98.5 & 0.2 $\rightarrow$ 0.88                                                                         \\ 
\hline
Diabetic Retinopathy (2014) & 0.49  & 81.8 & 1.46 & 82.3 & 0.5 $\rightarrow$ 0.97                                                                         \\ 
\hline
NSL KDD (5-class, 1999) & 11.52 & 88.5 & 15.27 & 90.1 & 1.6 $\rightarrow$ 3.75                                                                         \\ 
\hline
NSL KDD (2-class, 1999) & 13.22 & 84.4 & 16.78 & 85.2 & 0.8 $\rightarrow$ 3.56                                                                         \\ 
\hline
Breast cancer (1995)     & 18.58 & 99.3 & 23.86 & 100 & 0.7 $\rightarrow$ 5.28                                                                         \\ 
\hline
Sonar                & 4.18  & 88.1 & 5.70 & 90.5 & 2.4 $\rightarrow$ 1.52                                                                         \\ 
\hline
Wine (1991)              & 4.56  & 100 & 6.83 & 100 & 0.0 $\rightarrow$ 1.82                                                                         \\
\hline
\end{tabular}
}
\end{table}
Table \ref{tab:time-performance tradeoff} gives an idea about time-performance tradeoff after merging base models. The following could be incurred from the table: how much additional time was spent in order to enhance the performance of the base model. 
 \begin{figure}[!ht]
  \centering
\includegraphics[width=0.75\textwidth,height=12cm,keepaspectratio]{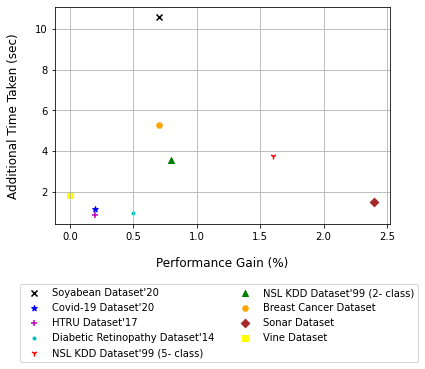}
  \caption{Trade-off between Time and Performance}
  \label{Fig: Res: Time perf tradeoff.png}
\end{figure}
Figure \ref{Fig: Res: Time perf tradeoff.png} aids in a clearer understanding of the balance between time and performance for each dataset. It can be observed that the additional time required to improve performance is absolutely minuscule. 
It is because we sought to use as few base models as possible. In real-world scenarios, it is a great compromise that makes the proposed approach incredibly adaptable. To summarize, the proposed technique is more comprehensible and enhances performance at a reasonable time tradeoff.
\end{enumerate}

\section{Conclusion}
\label{sec:cnf}
In this paper, we propose an efficient Human-Centered Ensemble Learning approach to Supervised Learning that can reduce overhead such as high cost, resources, time, and lack of understandability, which poses many challenges in real-world applications. The proposed method begins by identifying diverse strong base classifiers. We then employ a variety of strong Machine Learning and Deep Learning models on subsets of original datasets obtained through various feature selection and reduction techniques. Following that, we find a minimal subset of the obtained models by first grouping models using the hierarchical clustering concept and then selecting only the top-performing models from each cluster as the base learners for Ensemble Learning. This method ensures that the Ensemble Learning approach is more understandable while also reducing time, cost, and resource overhead, resulting in increased use of Stacking Ensembles in real-world applications. We test our hypothesis on nine real-world datasets. The experimental results show that using the above-mentioned approach, a minimal subset of classifiers outperforms individual classifiers. We also unearthed that diverse base classifiers have little or no correlation with performance. In addition, by comparing our results to those of recent research papers in this direction that used the same datasets, we show that this approach outperforms existing approaches.

\bibliography{08_bibliography}

\end{document}